\documentclass{elsart}

\usepackage{amssymb}
\usepackage[english]{babel}
\usepackage{latexsym}
\usepackage{graphicx}

\begin{document}
\begin{frontmatter}



\title{The Influence of Synthesis Parameters on Microstructures and Superconducting Properties of MgB$_{2}$ Bulks}


\author[label1,label2]{Y.~F.~Wu\corauthref{cor1}}
\corauth[cor1]{Corresponding author. Tel:+86-29-86231079;
fax:+86-29-86224487} \ead{wyf7777@tom.com}
\author[label1]{Y.~Feng}
\author[label1]{G.~Yan}
\author[label2]{J.~S.~Li}
\author[label1]{H.~P.~Tang}
\author[label1]{S.~K.~Chen}
\author[label3]{Y.~Zhao}
\author[label3]{M.~H.~Pu}
\author[label1]{H.~L.~Xu}
\author[label1]{C.~S.~Li}
\author[label1]{Y.~F.~Lu}

\address[label1]{Northwest Institute for Nonferrous Metal Research, P. O. Box 51, Xian, Shaanxi 710016, P. R. China}
\address[label2]{Northwestern Polytechnical University, Xi'an 710012, P.R.China}
\address[label3]{Southwest Jiaotong University, Si Chuan 610031, P.R.China}

\begin{abstract}
% Text of abstract

We succeeded in the synthesis of high-\textit{J$_{c}$} MgB$_{2}$
bulks via high energy ball-milling of elemental Mg and B powder at
ambient temperatures. Alternatively to long-time mechanical alloying
technique, the mixed powder was ball-milled for only 1h and the
completed reaction is achieved by subsequent annealing. The
correlations among synthesis parameters, microstructures and
superconducting properties in MgB$_{2}$ bulks were investigated.
Samples were characterized by XRD and SEM, and the magnetization
properties were examined in a SQUID magnetometer. The
high-\textit{J$_{c}$}, approximately
1.7$\times\,$10$^{6}$A/cm$^{2}$(15K, 0.59T), and improved flux
pinning were attribute to a large number of grain boundaries
provided by small grain size.
\end{abstract}


\end{frontmatter}

% main text
\section{Introduction}
\label{section:intro}
The improvement of the intrinsic properties of
MgB$_{2}$ was recognized as a decisive goal to enable potential
applications~\cite{Larbalestier:01}. Now, MgB$_{2}$ is showing
higher \textit{H$_{c2}$} than that of conventional NbTi and
Nb$_{3}$Sn and becoming the first promising metallic superconductor
applicable at ~20K. In particular, very high upper critical field,
\textit{H$_{c2}$}(0), exceeding 30T has been reported for MgB$_{2}$
wires, tapes and bulks. However, it was recognized that the
irreversibility field \textit{H$_{irr}$}(T) of the samples prepared
by standard solid state reaction is apparently lower than
\textit{H$_{c2}$} due to weak flux pinning. A typical relationship,
\textit{H$_{irr}$}(T)$\sim$0.5\textit{H$_{c2}$}(T), has been
observed for the undoped MgB$_{2}$
bulks~\cite{Yamamoto:05,Gumbel:02}. It was also shown that for
MgB$_{2}$, contrast to high \textit{T$_{c}$} superconductors, grain
boundaries in MgB$_{2}$ are not acting as impediments for
superconducting currents~\cite{Larbalestier:01,Bugoslavsky:01}.
Unfortunately, a rapid drop of \textit{J$_{c}$} at high fields,
probably related to weak-link-like behavior, can be seen in most
studies
~\cite{Giunchi:03,Civale:02,Suo:01,Dou:02,Fujii:02,Flukiger:03}. The
predominant pinning mechanism in MgB$_{2}$ is still controversy.
However, a microstructure with very small-sized defects (e.g. grain
boundary) should be favorable for the optimum pinning of magnetic
flux lines. The high energy ball-milling technique facilitates the
formation of an optimal microstructure with small particle sizes.
Hence, a high grain boundary density is obtained, which is expected
to enhance the magnetic flux pinning ability and to improve the
critical currents in external magnetic fields.

\section{Experimental}
\label{section:Experimental}

In this study Mg(99.8\%) and amorphous B(95+\%) powders with 5wt\%
Mg surplus were filled under purified Ar-atmosphere into an agate
milling container and milling media. The milling was performed on a
SPEX 8000M mill for 1h using a ball-to-powder mass ratio of 3. The
milled powders were then cold pressed to form pellets with a
diameter of 20mm and a height of 3mm. The pellets were placed in an
alumina crucible inside a tube furnace under ultra-high purity
Ar-atmosphere. The samples were heated at different temperatures for
different times, then cooled down to the ambient temperature.

A Quantum Design SQUID magnetometer was used to measure the AC
magnetic susceptibility of the samples over a temperature range of 5
to 50 K under an applied field of 1Oe. Magnetization versus magnetic
field (M-H) curves were also measured on rectangular-shaped samples
at temperatures of 10 and 15K under magnetic fields up to 90000Oe to
determine the critical current density \textit{J$_{c}$}(\textit{H}).

The phase compositions of the samples were characterized by the
APD1700 X-ray diffraction instrument. The surface morphology and
microstructures of the samples were characterized by the JSM-6460
and the JSM-6700F scanning electron microscope.
% \appendix

% \section{}
% \label{}

\section{Results and discussion}
\label{section3}
\subsection{The influence of sintering temperature to the
phase composition, microscopy and superconducting properties of bulk
MgB$_{2}$}
\label{section3.1}

The X-ray diffraction patterns of MgB$_{2}$ bulks sintered at
different temperatures are shown in figure 1. For the samples
sintered at $650\,^{\circ}\mathrm{C}$, $700\,^{\circ}\mathrm{C}$ and
$750\,^{\circ}\mathrm{C}$, almost single phase MgB$_{2}$ appears in
the patterns with minor fraction of MgO. The percentage of MgO is
apparently increased for the sample sintered at
$800\,^{\circ}\mathrm{C}$, indicating easy oxidation of Mg at higher
temperature.

Shown in Fig.2(a) is the surface morphology of the sample sintered
at $650\,^{\circ}\mathrm{C}$. Well-developed coarse columnar grains
can be seen in the figure. Evidently, it could not connect very well
in the surface. There are also some black impurity phases for the
sample sintered at $650\,^{\circ}\mathrm{C}$,see Fig.2(b). The
energy spectrum analysis showed that the black one is the B-rich
phase, which is due to the stability of the higher borides at lower
temperature, see Fig.3. In contrast, the grains of the samples
sintered at $700\,^{\circ}\mathrm{C}$ and $750\,^{\circ}\mathrm{C}$
are equiaxial, fine and well-connected,shown in Fig.2(c)and
Fig.2(e). Additionly, there are only a few impurity phases for these
samples, see Fig.2(d) and Fig.2(f). It exhibits the sintering
temperatures are appropriate right here. A broad grain size
distribution can be seen for the sample sintered at
$800\,^{\circ}\mathrm{C}$, as shown in Fig.2(g). The large number of
existing impurity phases,see Fig.2(h) would obviously take bad
effect to the grain connectivity and lower the superconducting
properties of the sample. It is clearly not the suitable temperature
for preparation of bulk MgB$_{2}$.

Fig.4 shows the AC magnetic susceptibility as a function of the
temperature for the samples sintered at different temperatures. A
constant magnetic field of 1Oe was applied. As we can see, all
samples have sharp transitions. But it is interesting that the
superconducting transition of the samples sintered at
$700\,^{\circ}\mathrm{C}$, $750\,^{\circ}\mathrm{C}$ and
$800\,^{\circ}\mathrm{C}$ ( \textit{T$_{c}$}$\sim$1.5 K) is about
2.5 times sharper than that of the samples sintered at
$650\,^{\circ}\mathrm{C}$ ( \textit{T$_{c}$}$\sim$4 K). It is not
clear what caused the transition broaden. However, the broadening
could be related to microstructural changes induced by the heat
treatment temperature. It seems that the columnar grains lead to
worse grain connectivity than the impurity phases.

The magnetization curves (M-H) for the samples sintered at different
temperatures are shown in Fig.5. \textit{J$_{c}$}(H) was determined
by Bean critical state model, as shown in Fig.6. The
\textit{J$_{c}$} at 0K is nearly equal for all the samples. However,
the sample sintered at $750\,^{\circ}\mathrm{C}$ has a significantly
higher \textit{J$_{c}$} than the other samples in magnetic field,
indicating improved magnetic flux pinning of it. The differences in
\textit{J$_{c}$} among them increase with field. The sample sintered
at $650\,^{\circ}\mathrm{C}$ shows a steep drop in \textit{J$_{c}$}
at higher fields (H$>$4T and T = 10K). It is most probably because
of the columnar grains existing in the sample lead to bad grain
connectivity, which severely limits its \textit{J$_{c}$}
performance. No such steep drop in \textit{J$_{c}$} is observed for
the other samples.

\subsection{The influence of the holding time to the phase
composition, microscopy and superconducting properties of bulk
MgB$_{2}$}

The X-ray diffraction patterns of bulk MgB$_{2}$ sintered for
different times are shown in figure 7. Mainly peaks of MgB$_{2}$ are
visible but some peaks of MgO are present. The differences in the
percentage of MgO phase are difficult to perceive in the patterns.

Shown in Fig.8(a) is the surface morphology of the samples sintered
for 0.5h. As we can see, the reaction between Mg and B is
incomplete. The big grains could not disappear completely.
Meanwhile, a large number of small grains have become visible in the
grain boundaries. It reveals that the grains are not well connected
in the surface for the sample sintered for short time. Some dark
grey and light grey impurity phases can be seen for the sample
sintered for 0.5h,see Fig.8(b). The energy spectrum analysis showed
that the dark one is the B-rich phase (Fig.9) and the light one is
the magnetism oxidation compound (Fig.10). However, the grains of
the sample sintered for 1h are equiaxial, fine and
well-connected,see Fig.8(c). It only has a few small-sized impurity
phases,see Fig.8(d). Although the grains are fine as well for the
sample sintered for 3h, see Fig.8(e), a great many existing
impurities would apparently deteriorate its \textit{J$_{c}$}
performance as shown in Fig.8(f). It indicates undue shorter or
longer holding time lead to much more impurity phases and inferior
microstructures.

Shown in Fig.11 are magnetization-field loops for samples sintered
for different times. \textit{J$_{c}$}(H) was deduced from the
hysteresis loops using the Bean model, see Fig. 12. The sample
sintered for 1h has a significantly higher \textit{J$_{c}$} and than
other samples in the magnetic field due to its optimum
microstructure. The \textit{H$_{irr}$} value is about 6.3T at 15K,
as determined from the closure of hyseresis loops with a criterion
of \textit{J$_{c}$}=10$^{2}$A/cm$^{2}$. The differences in
\textit{J$_{c}$} among them increase with field. The sample sintered
for 0.5 and 3h shows a steep drop in \textit{J$_{c}$} at higher
fields (H$>$4T and T = 15K). It indicates both columnar grains and
excessive impurity phases lead to the bad grain connectivity and
weak flux pinning, especially in high magnetic field.

\section{Summary}
\label{section:summary} In conclusion, the high-\textit{J$_{c}$}
MgB$_{2}$ bulks were prepared by short-time high energy ball
milling. The introduction of impurities (especially oxygen) during
the ball-milling process was minimized. A close relation among
microstructure, impurities and superconducting properties with the
synthesis parameters was detected. The irreversibility field of the
optimum sample reaches 6.3T at 15K and \textit{J$_{c}$} is high
around 1.7$\times\,$10$^{6}$A/cm$^{2}$(15K, 0.59T) at 15K in 0.59T.
The improved pinning of this material seems to be caused by enhanced
grain boundary pinning provided by the large number of grain
boundaries in the sample. Another crucial point during preparation
of MgB$_{2}$ is to avoid weak-link-like behavior at grain boundaries
due to a non-superconducting surface layer of magnetism oxidation or
B-rich compounds.

\section{ACKNOWLEDGMENT}
\label{section:Acknowledgment} This work was partially supported by
National Natural Science Foundation project(grant 50472099)and
National 973 project(grant 2006CB601004).

\newpage
\textbf{Figure captions}

Fig.1 XRD patterns of MgB$_{2}$ bulks sintered at (a)$650
\,^{\circ}\mathrm{C}$, (b) $700\,^{\circ}\mathrm{C}$,
(c)$750\,^{\circ}\mathrm{C}$ and (d) $800\,^{\circ}\mathrm{C}$.
Peaks of MgB$_{2}$ and MgO are marked by solid circles and squares,
respectively.

Fig.2 The SEM photograph of the MgB$_{2}$ bulks sintered at
different temperatures:
(a)$650\,^{\circ}\mathrm{C}$$\times\,$15,000
(b)$650\,^{\circ}\mathrm{C}$$\times\,$4,000
(c)$700\,^{\circ}\mathrm{C}$$\times\,$15,000
(d)$700\,^{\circ}\mathrm{C}$$\times\,$4,000
(e)$750\,^{\circ}\mathrm{C}$$\times\,$15,000
(f)$750\,^{\circ}\mathrm{C}$$\times\,$4,000
(g)$800\,^{\circ}\mathrm{C}$$\times\,$15,000 and
(h)$800\,^{\circ}\mathrm{C}$$\times\,$4,000.

Fig.3 The EDX analysis of the second phase in the sample sintered at
$650\,^{\circ}\mathrm{C}$.

Fig.4 The AC Magnetic susceptibility as a function of temperature of
MgB$_{2}$ bulks. A constant magnetic field of 1Oe was applied.

Fig.5 Magnetization \textit{M} as a function of magnetic field
\textit{H} at 10K for the samples sintered at different
temperatures.

Fig.6 Magnetization critical current density \textit{J$_{c}$} as a
function of magnetic field \textit{H} at 10K for the samples
sintered at different temperatures.

Fig.7 The XRD patterns of MgB$_{2}$ sintered at
$750\,^{\circ}\mathrm{C}$ for different holding times.Peaks of
MgB$_{2}$ and MgO are marked by solid circles and squares,
respectively.

Fig.8 The SEM photograph of the MgB$_{2}$ bulks sintered for
different holding
times:(a)0.5h$\times\,$15,000(b)0.5h$\times\,$4,000
(c)1h$\times\,$15,000(d)1h$\times\,$4,000 (e)3h$\times\,$15,000 and
(f)3h$\times\,$4,000.

Fig. 9 The EDX analysis of dark gray second phase in the 0.5h
sintered sample.

Fig.10 The EDX analysis of light gray second phase in the 0.5h
sintered sample.

Fig.11 Magnetization \textit{M} as a function of magnetic field
\textit{H} at 15K for samples sintered for different holding times.

Fig.12 Magnetization critical current density \textit{J$_{c}$} as a
function of magnetic field H at 15 K for the samples sintered for
different holding times.

\newpage
Fig.1
%\section{XRD Analysis for diff temp}
%%%%%%%%%%%%%%%%%%%%%%%%  FIGURE 1  XRD Analysis for diff temp  %%%%%%%%%%%%%%%%%%%%%%%%%
\begin{figure}[hp]
\centering
\includegraphics[width=200pt]{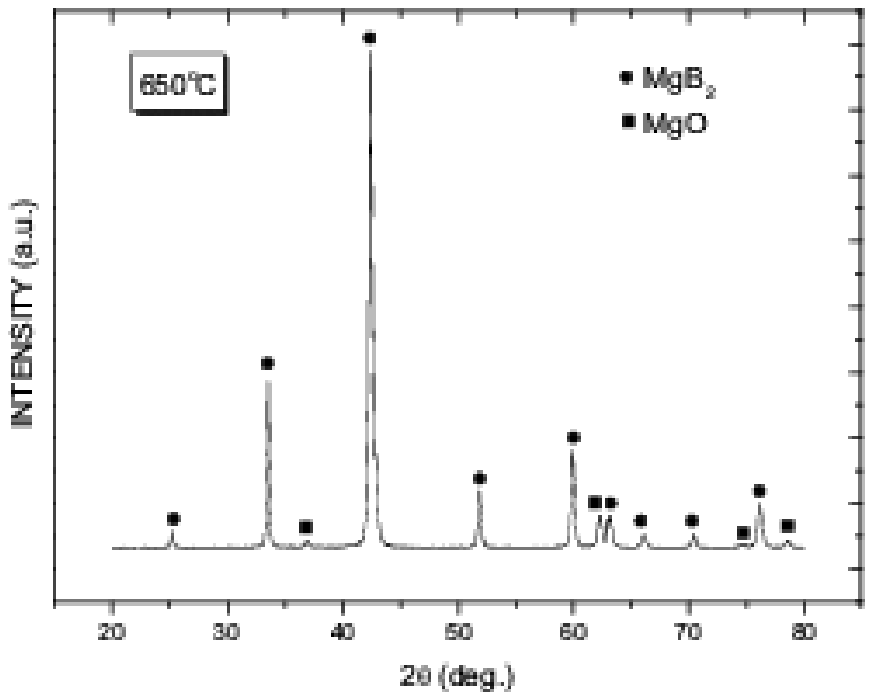}%
\includegraphics[width=200pt]{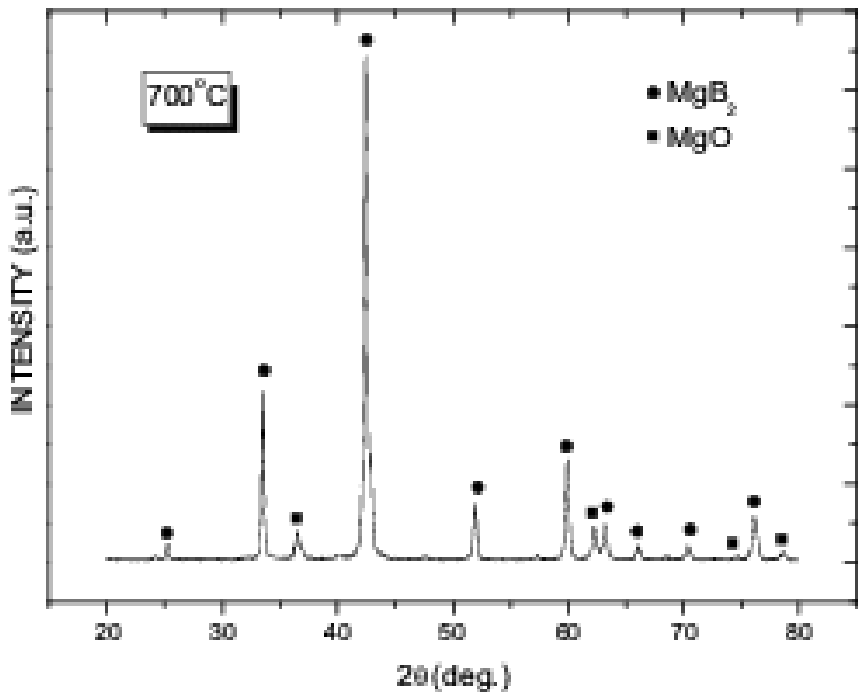}
\includegraphics[width=200pt]{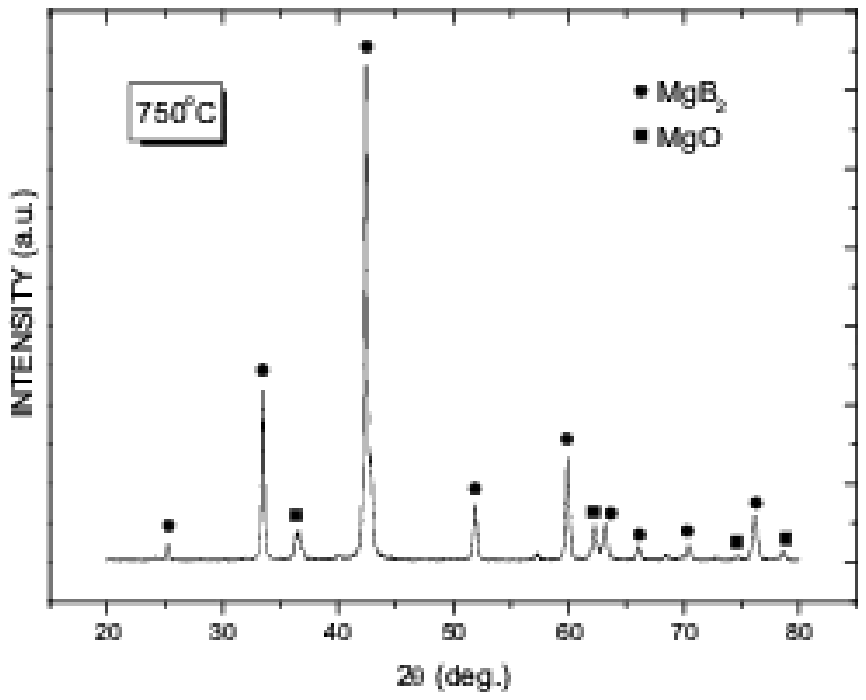}%
\includegraphics[width=200pt]{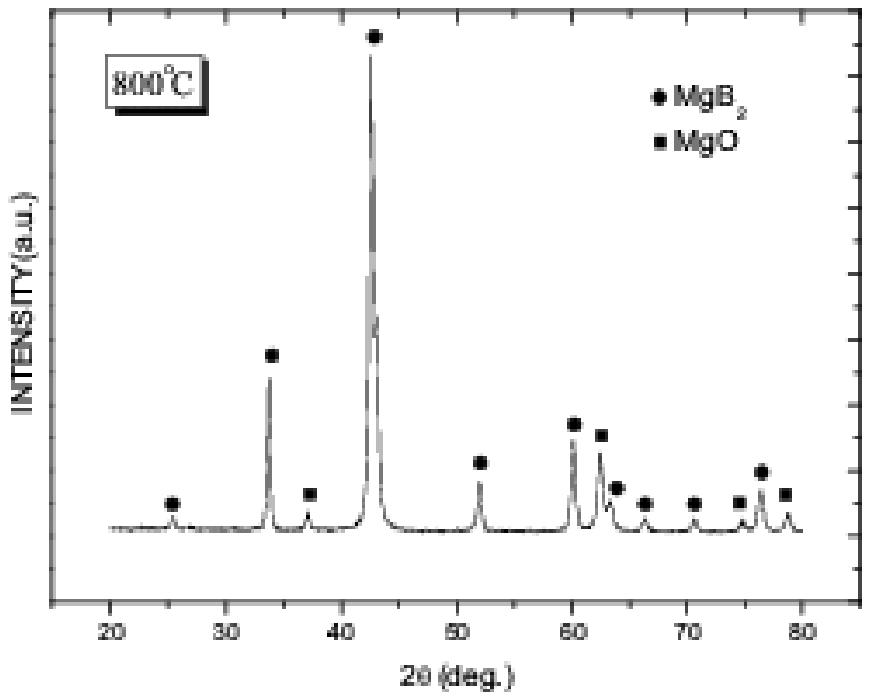}

\end{figure}
%%%%%%%%%%%%%%%%%%%%%%%%%%%%%%%%%%%%%%%%%%%%%%%%%%%%%%%%%%%%

\newpage
Fig.2
%\section{SEM for diff temp}
%%%%%%%%%%%%%%%%%%%%%%%%  FIGURE 2 SEM for diff temp  %%%%%%%%%%%%%%%%%%%%%%%%%
\begin{figure}[hp]
\centering
\includegraphics[width=180pt]{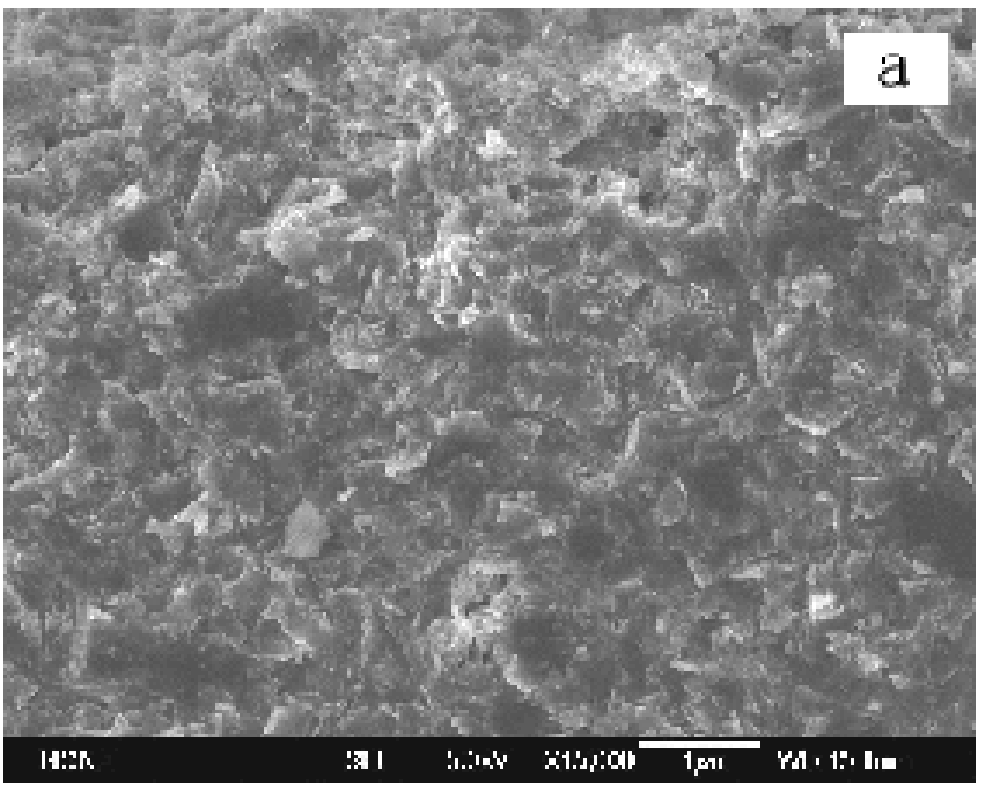}%
\includegraphics[width=180pt]{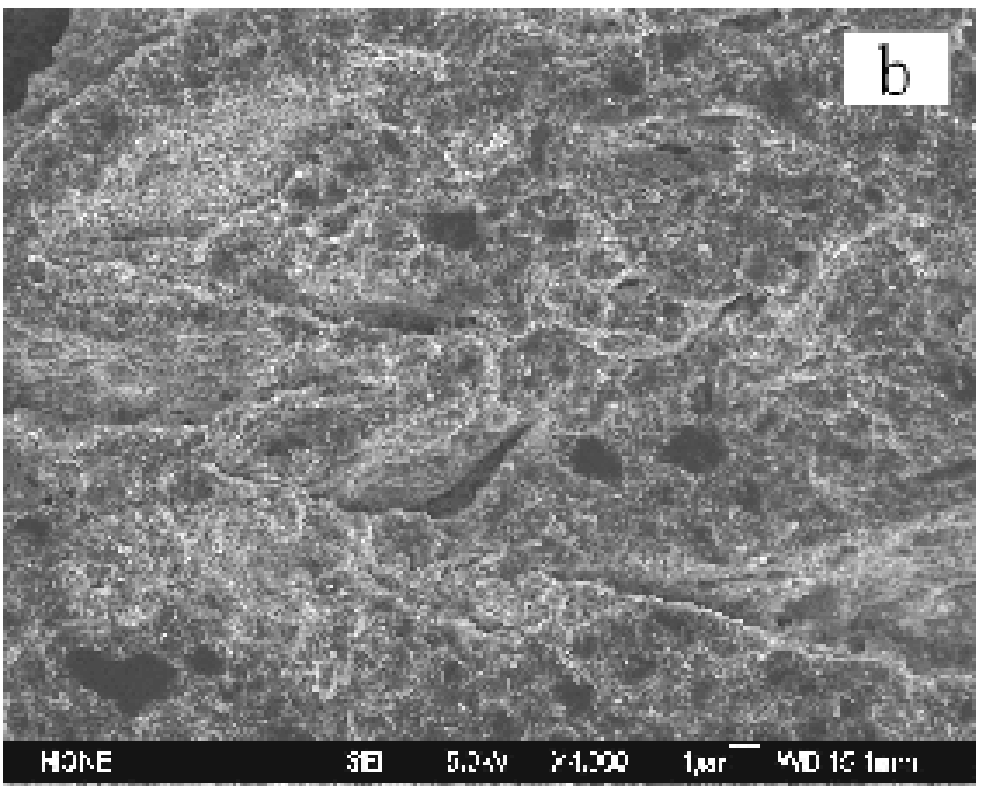}
\includegraphics[width=180pt]{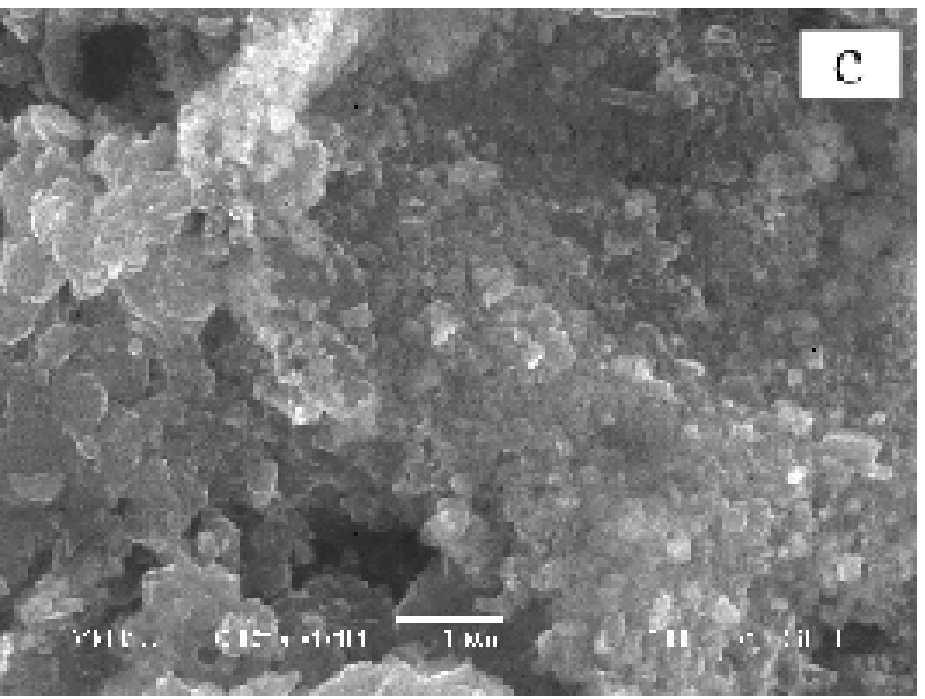}%
\includegraphics[width=180pt,height=135pt]{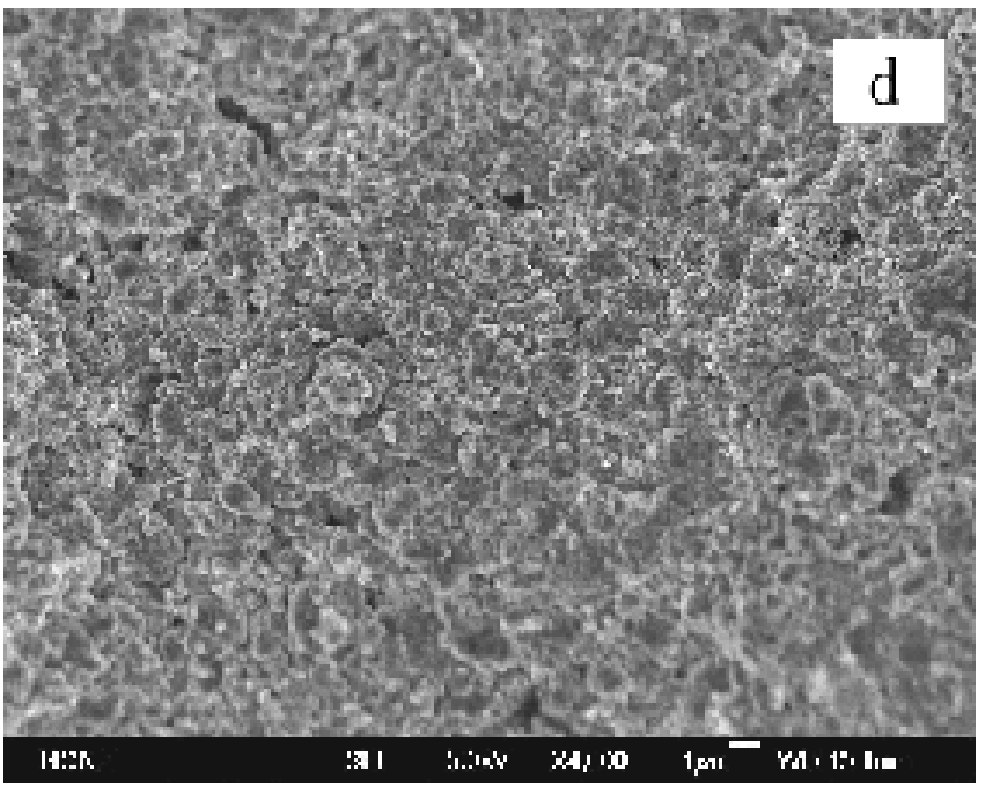}
\includegraphics[width=180pt]{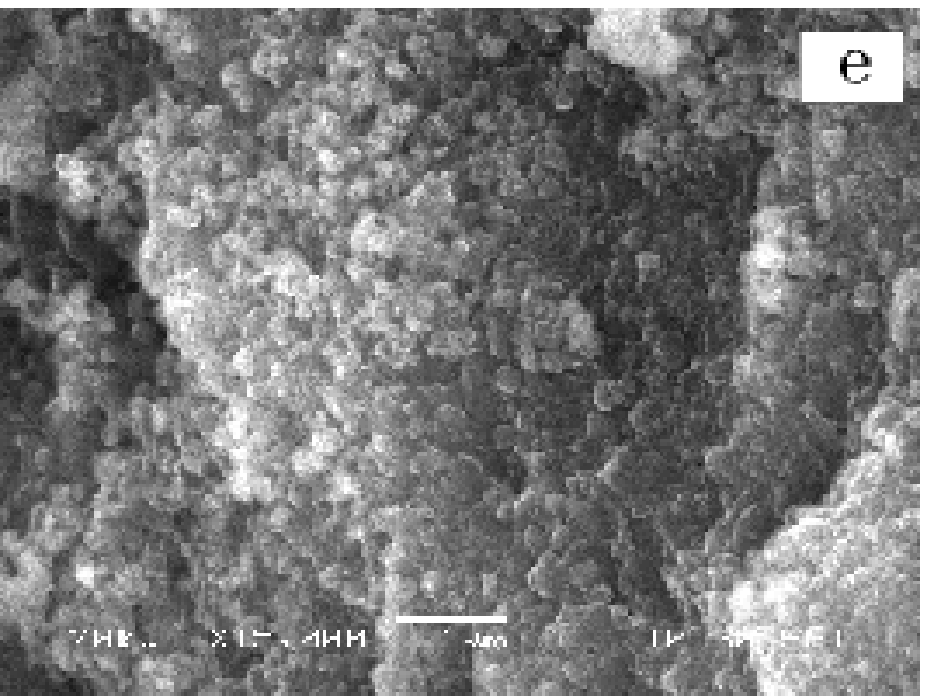}%
\includegraphics[width=180pt,height=135pt]{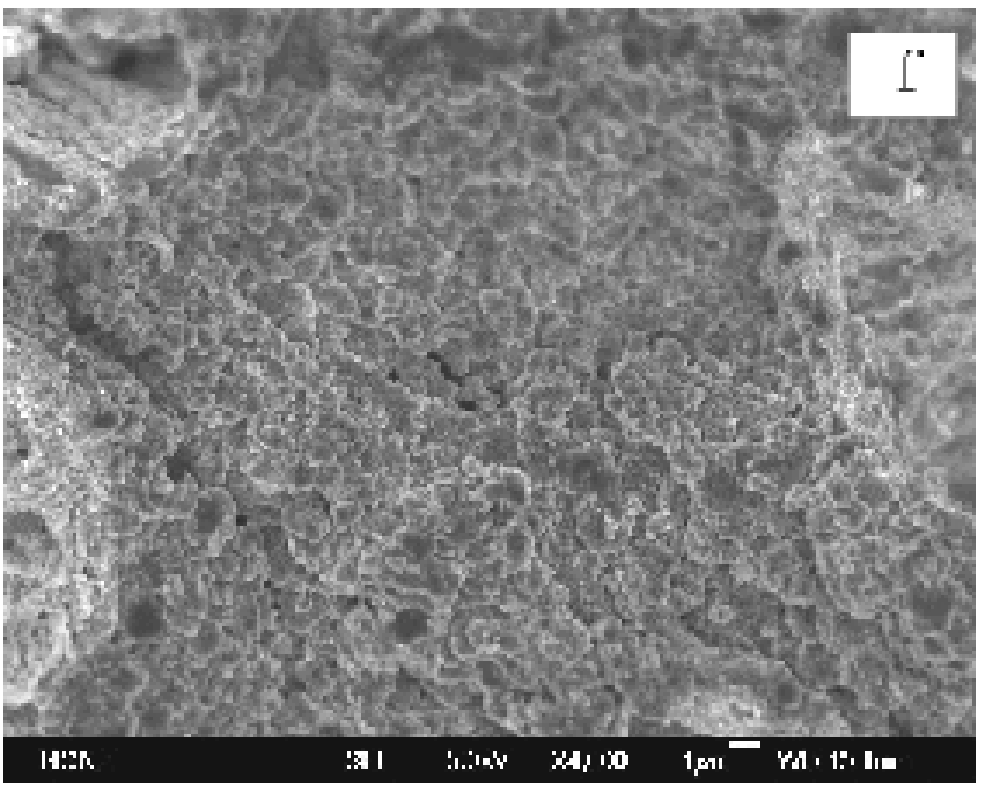}
\includegraphics[width=180pt]{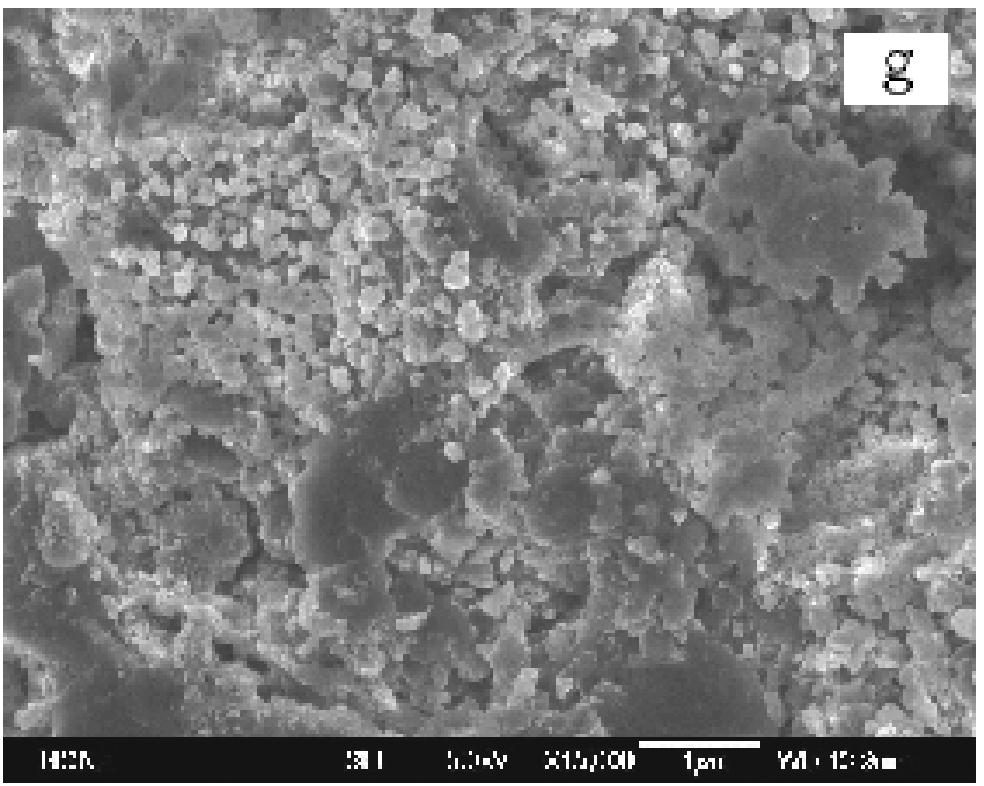}%
\includegraphics[width=180pt]{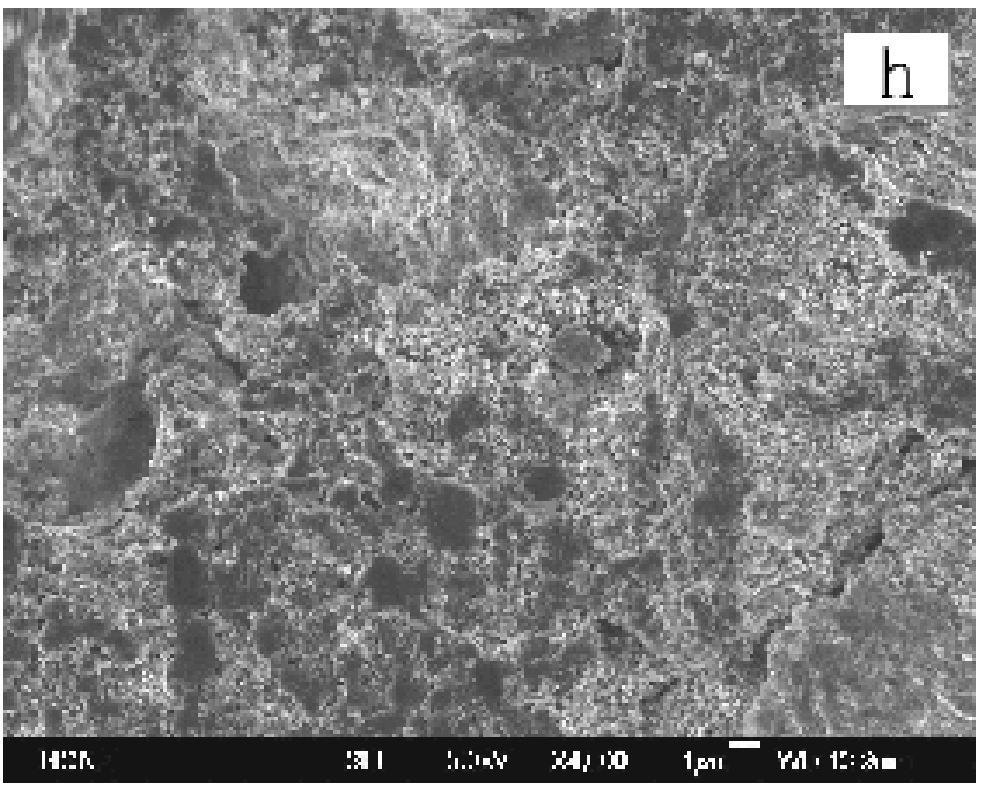}

\end{figure}

%%%%%%%%%%%%%%%%%%%%%%%%%%%%%%%%%%%%%%%%%%%%%%%%%%%%%%%%%%%%

\newpage
Fig.3
%\section{EDX second phase 650}
%%%%%%%%%%%%%%%%%%%%%%%%  FIGURE 3 EDX second phase 650  %%%%%%%%%%%%%%%%%%%%%%%%%
\begin{figure}[hp]
\centering
\includegraphics[width=180pt]{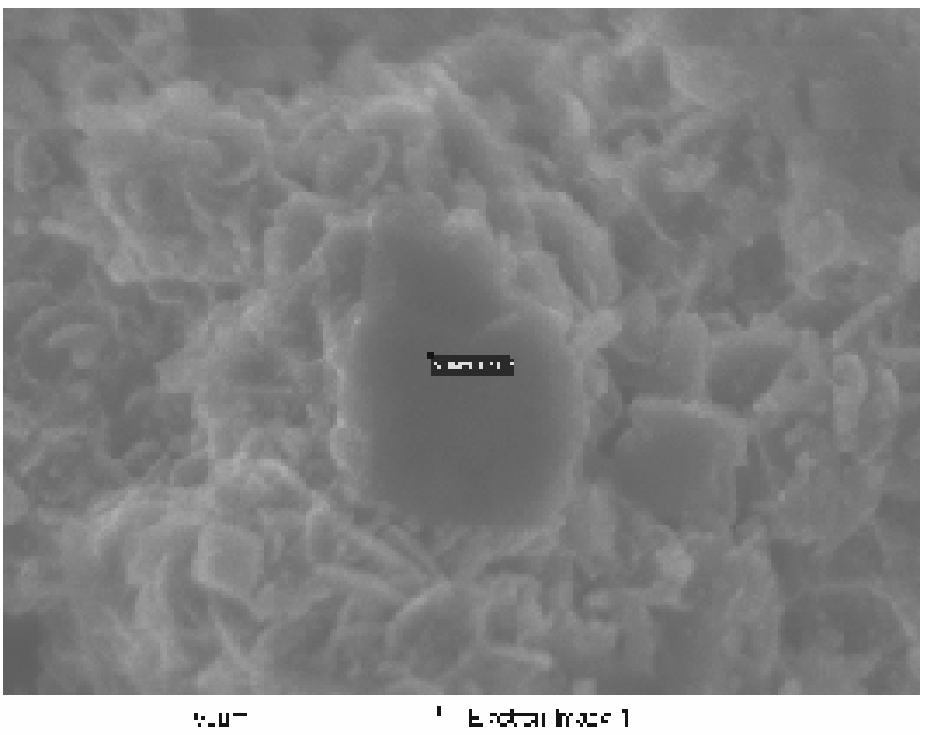}

\includegraphics[width=180pt]{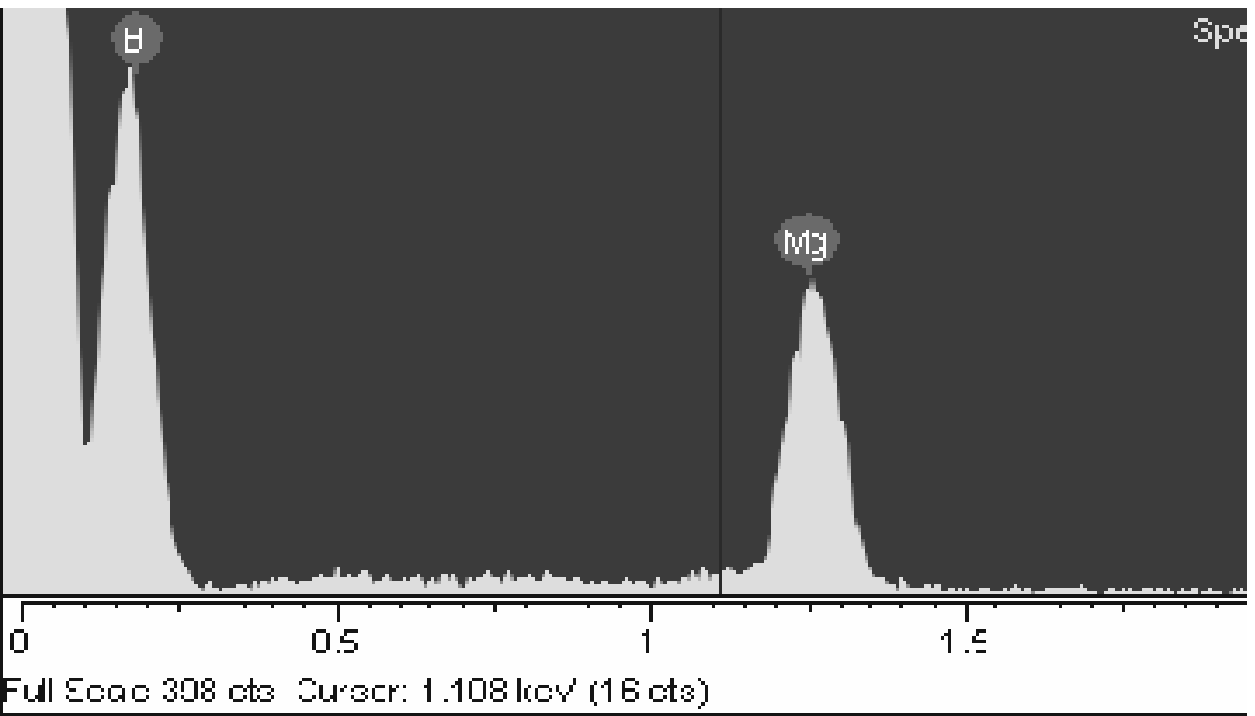}

\includegraphics[width=180pt]{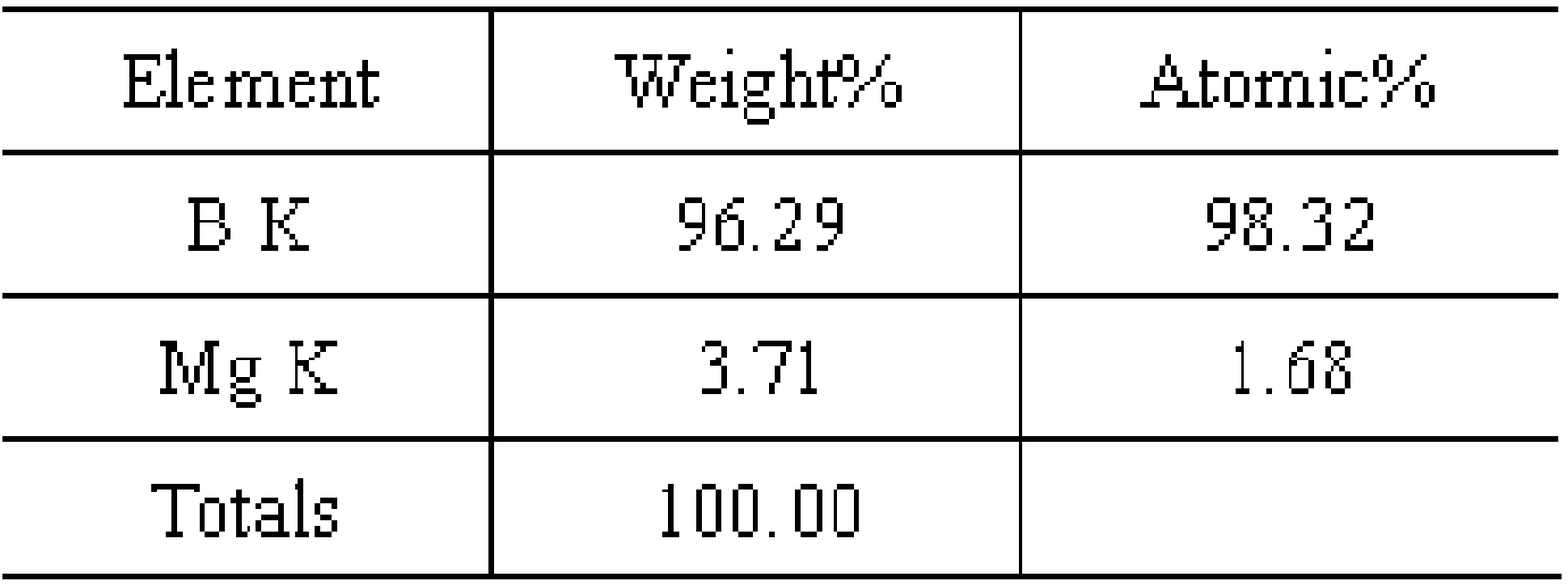}
\end{figure}

%%%%%%%%%%%%%%%%%%%%%%%%%%%%%%%%%%%%%%%%%%%%%%%%%%%%%%%%%%%%

\newpage
Fig.4
%\section{diff temp Tc}
%%%%%%%%%%%%%%%%%%%%%%%%  FIGURE 4 diff temp Tc  %%%%%%%%%%%%%%%%%%%%%%%%%
\begin{figure}[hp]
\centering
\includegraphics[width=300pt]{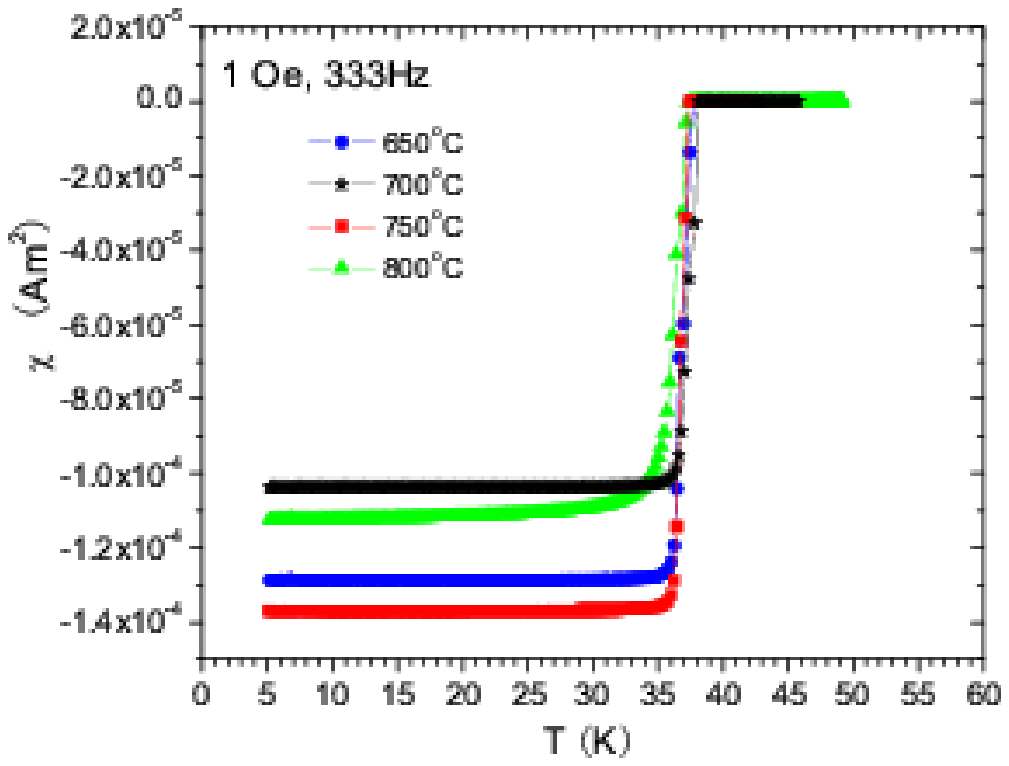}
\end{figure}

%%%%%%%%%%%%%%%%%%%%%%%%%%%%%%%%%%%%%%%%%%%%%%%%%%%%%%%%%%%%

\clearpage Fig.5
%\section{diff temp M-H}
%%%%%%%%%%%%%%%%%%%%%%%%  FIGURE 5 diff temp M-H  %%%%%%%%%%%%%%%%%%%%%%%%%
\begin{figure}[hp]
\centering
\includegraphics[width=300pt]{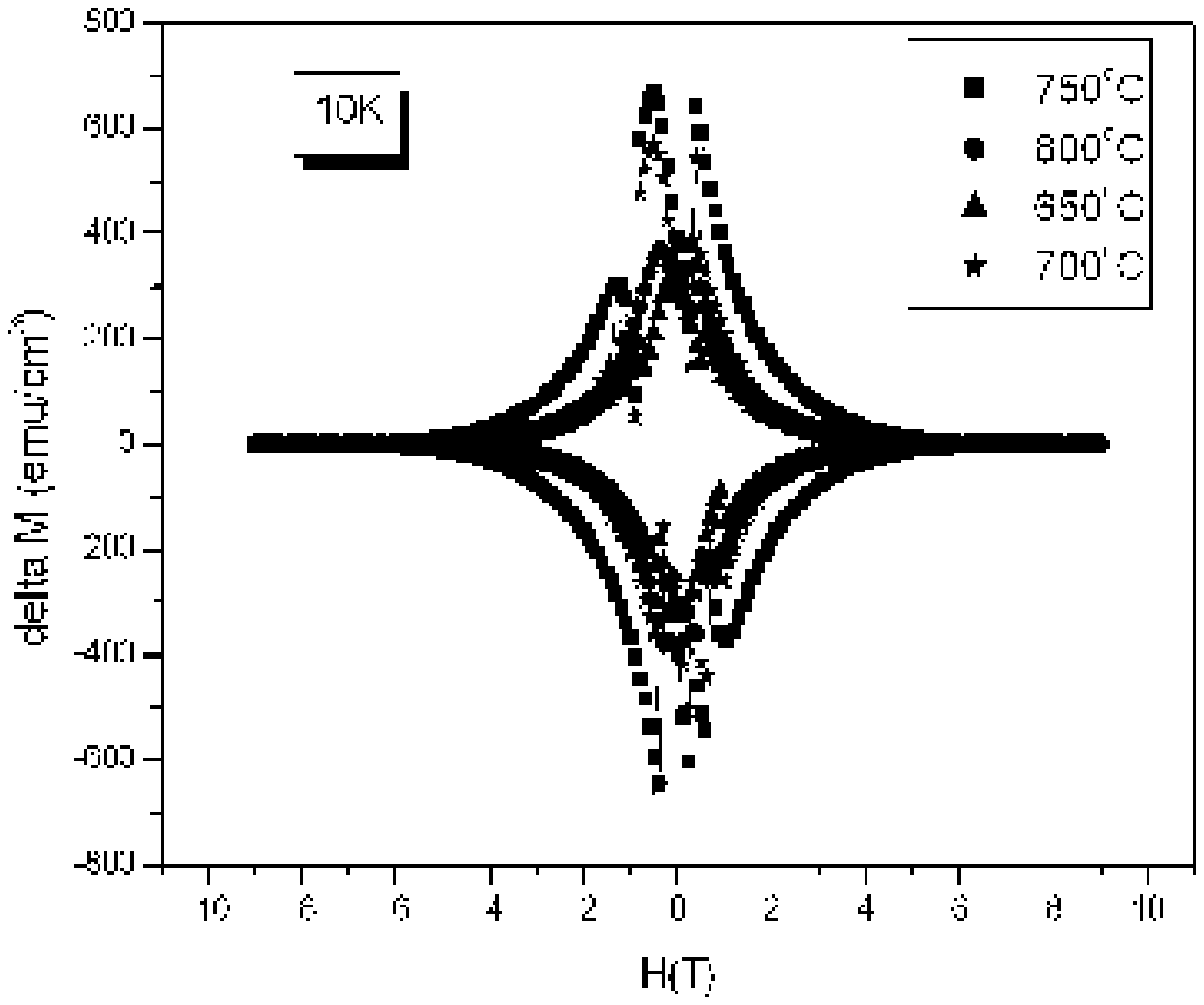}
\end{figure}

%%%%%%%%%%%%%%%%%%%%%%%%%%%%%%%%%%%%%%%%%%%%%%%%%%%%%%%%%%%%

\clearpage Fig.6
%\section{diff temp Jc-H}
%%%%%%%%%%%%%%%%%%%%%%%%  FIGURE 6 diff temp Jc-H  %%%%%%%%%%%%%%%%%%%%%%%%%
\begin{figure}[hp]
\centering
\includegraphics[width=300pt]{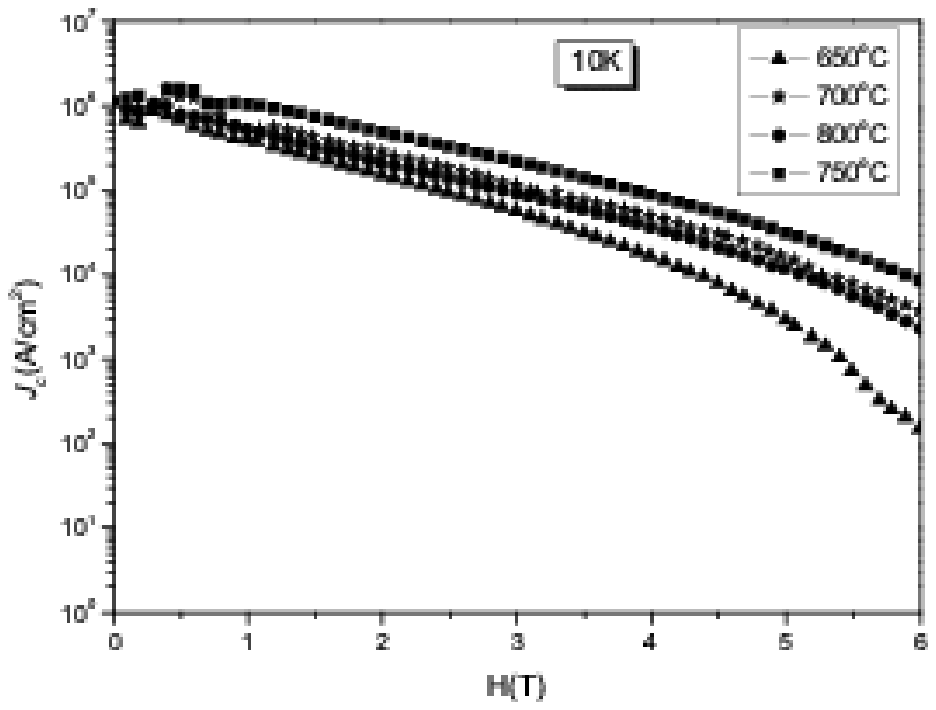}
\end{figure}

%%%%%%%%%%%%%%%%%%%%%%%%%%%%%%%%%%%%%%%%%%%%%%%%%%%%%%%%%%%%

\clearpage Fig.7
%\section{diff sintering time XRD}
%%%%%%%%%%%%%%%%%%%%%%%%  FIGURE 7 diff sintering time XRD %%%%%%%%%%%%%%%%%%%%%%%%%
\begin{figure}[hp]
\centering
\includegraphics[width=250pt]{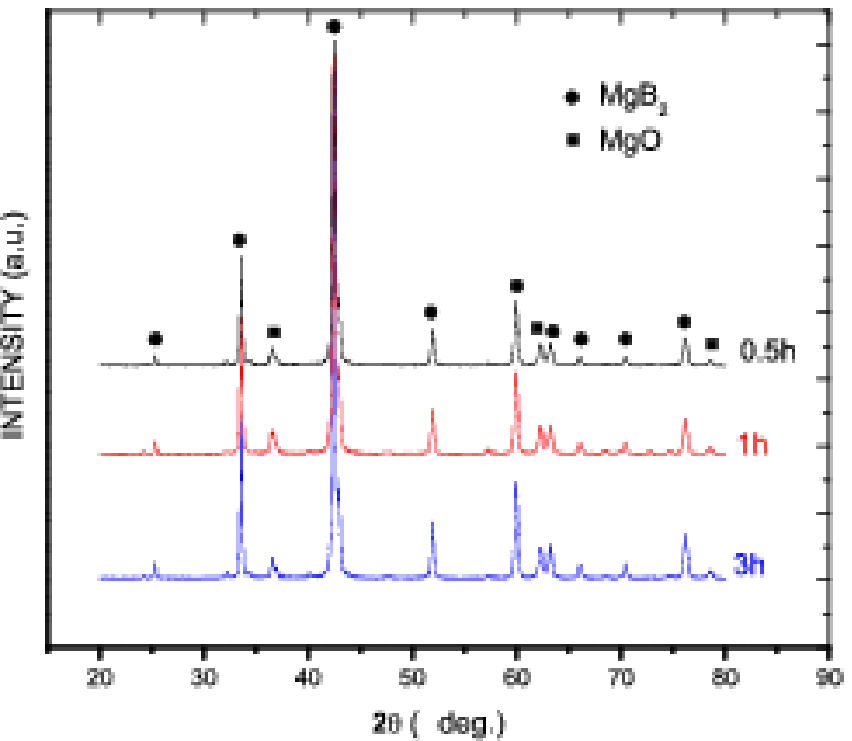}
\end{figure}

%%%%%%%%%%%%%%%%%%%%%%%%%%%%%%%%%%%%%%%%%%%%%%%%%%%%%%%%%%%%

\clearpage Fig.8
%\section{diff sintering time SEM}
%%%%%%%%%%%%%%%%%%%%%%%%  FIGURE 8 diff sintering time  SEM %%%%%%%%%%%%%%%%%%%%%%%%%
\begin{figure}[hp]
\centering
\includegraphics[width=180pt]{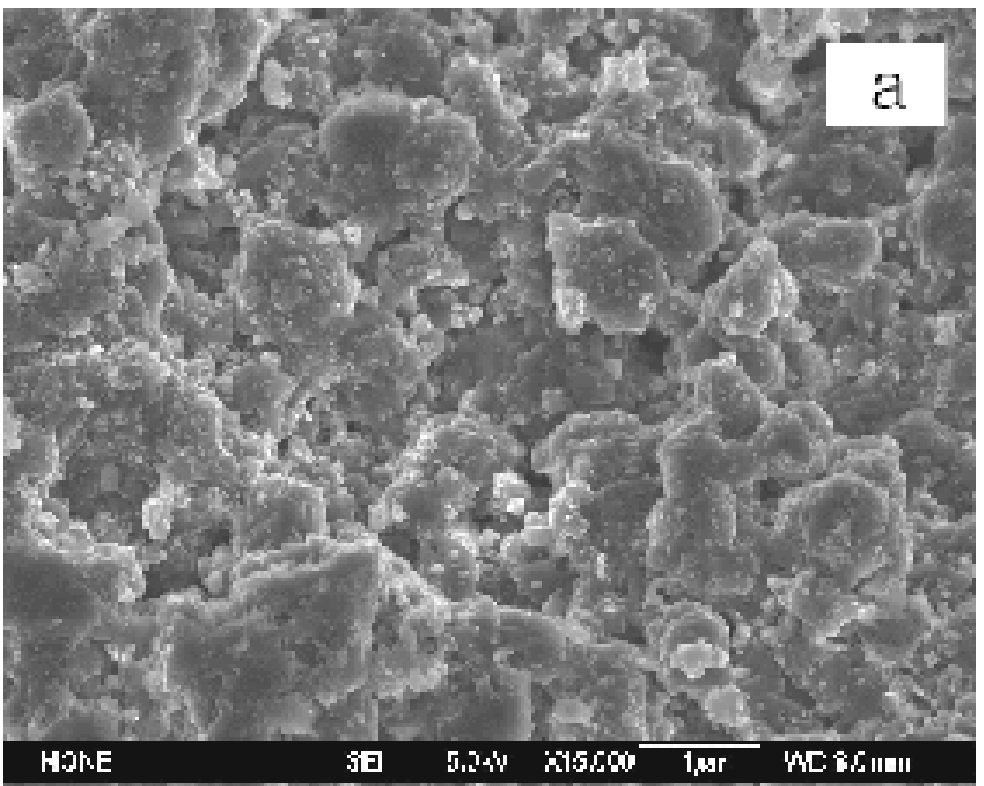}%
\includegraphics[width=180pt]{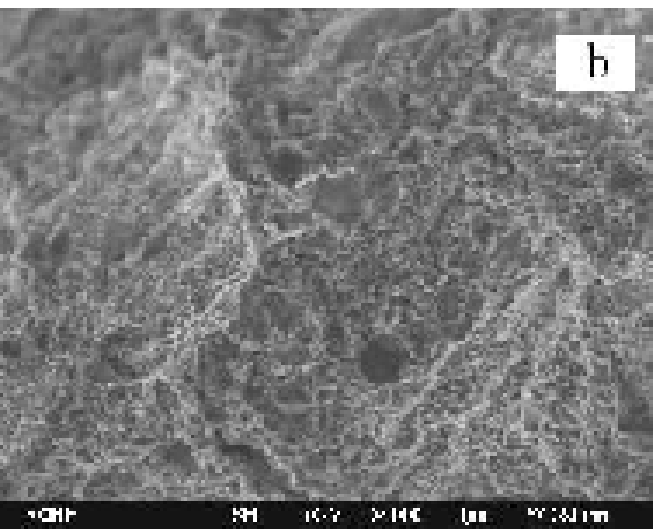}
\includegraphics[width=180pt]{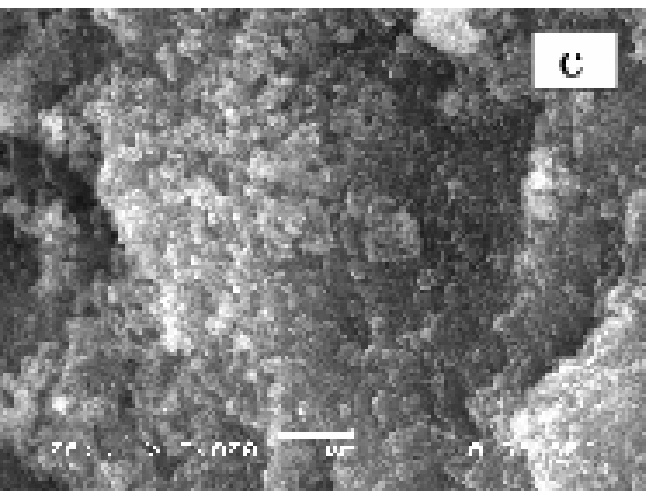}%
\includegraphics[width=180pt,height=135pt]{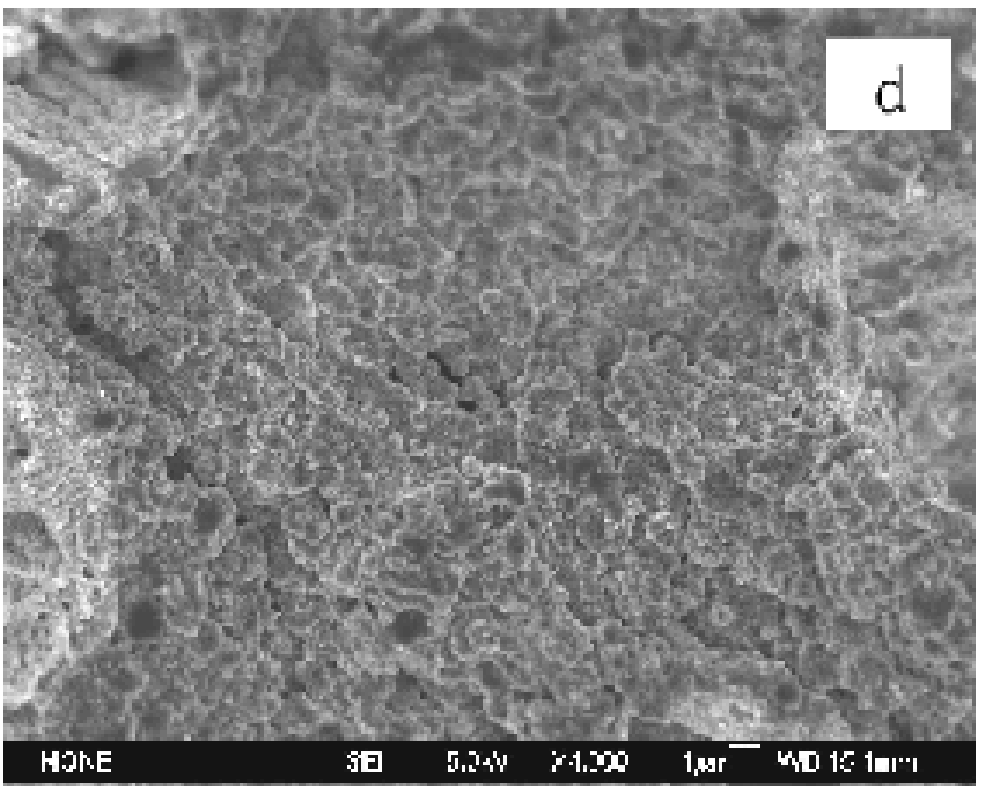}
\includegraphics[width=180pt]{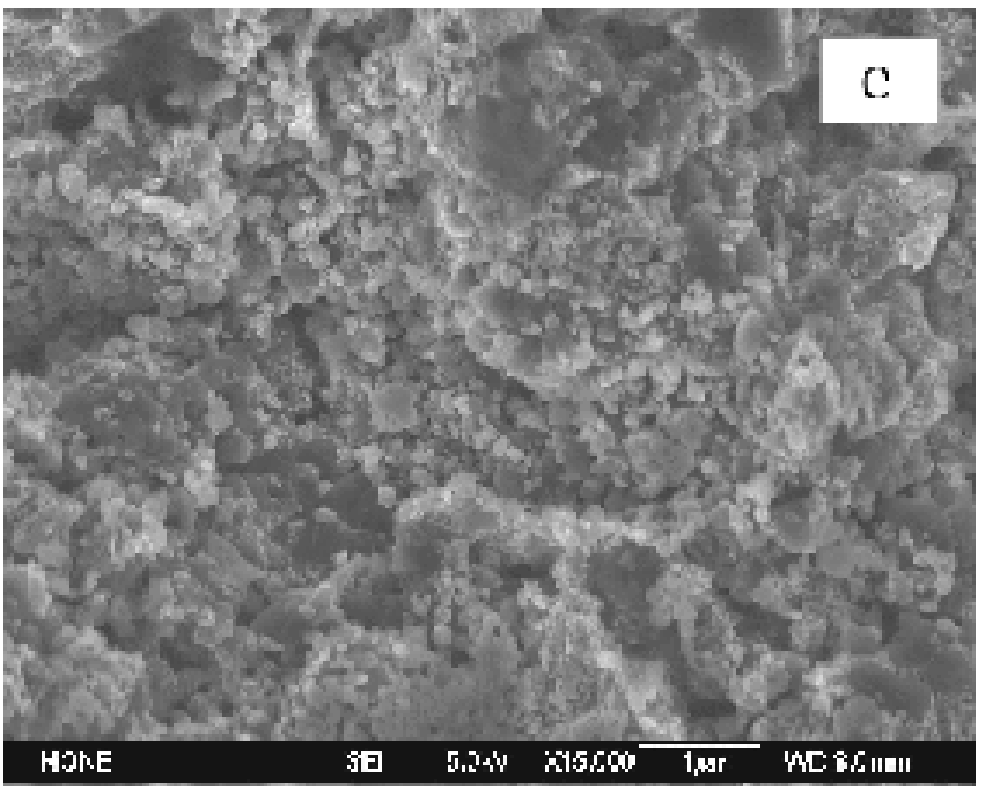}%
\includegraphics[width=180pt]{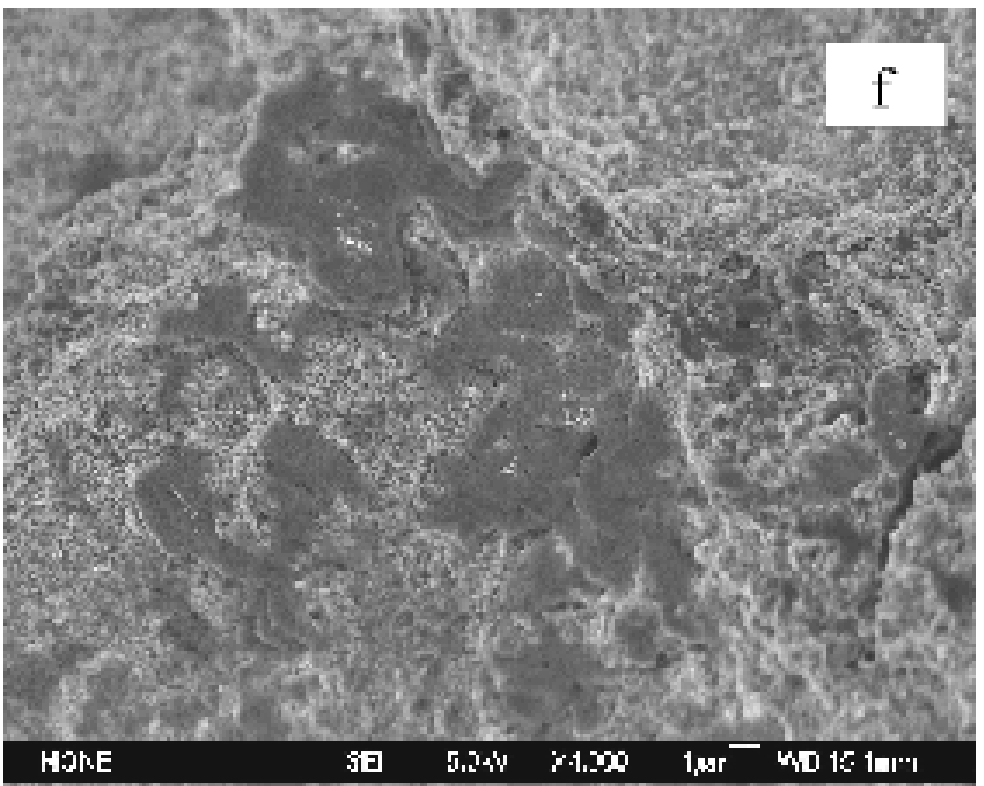}
\end{figure}

%%%%%%%%%%%%%%%%%%%%%%%%%%%%%%%%%%%%%%%%%%%%%%%%%%%%%%%%%%%%

\clearpage Fig.9
%\section{EDX DARK second phase 0.5h}
%%%%%%%%%%%%%%%%%%%%%%%%  FIGURE 9 EDX DARK second phase 0.5h  %%%%%%%%%%%%%%%%%%%%%%%%%
\begin{figure}[hp]
\centering
\includegraphics[width=180pt]{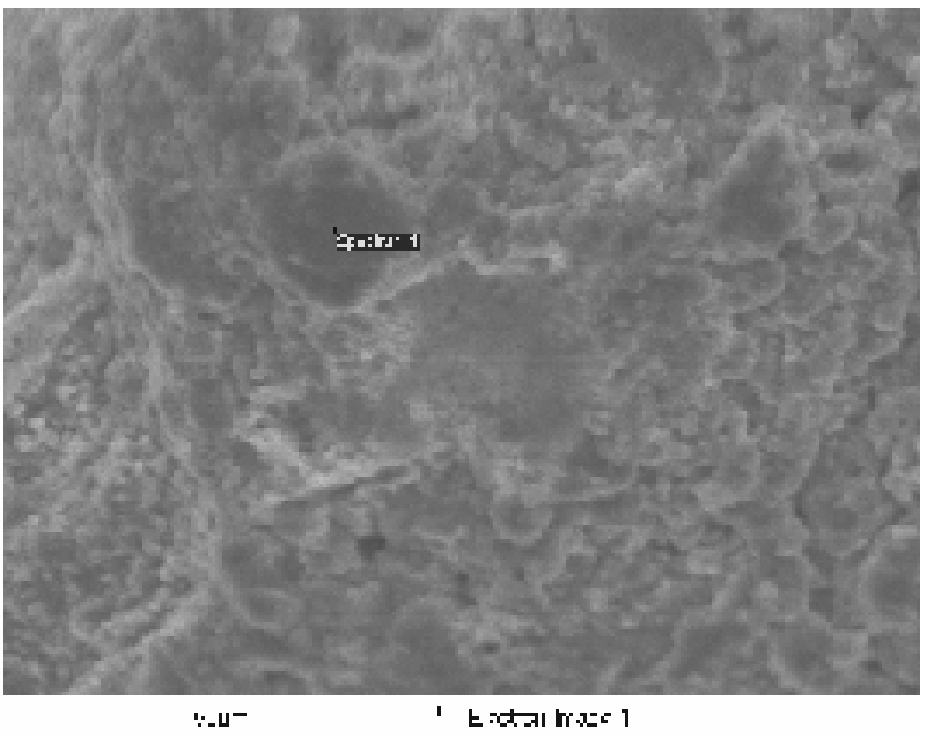}

\includegraphics[width=180pt]{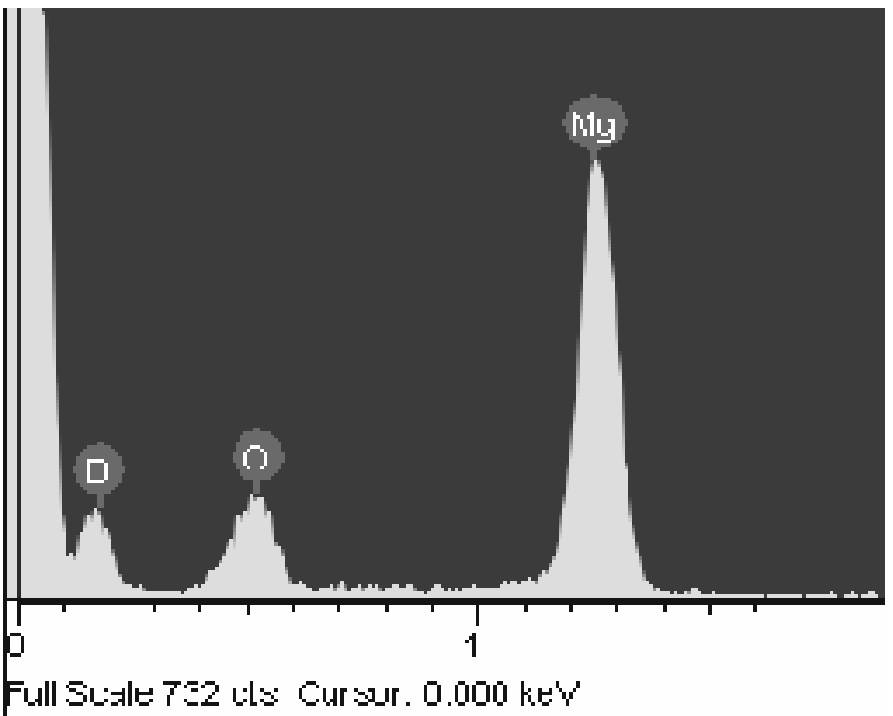}

\includegraphics[width=180pt]{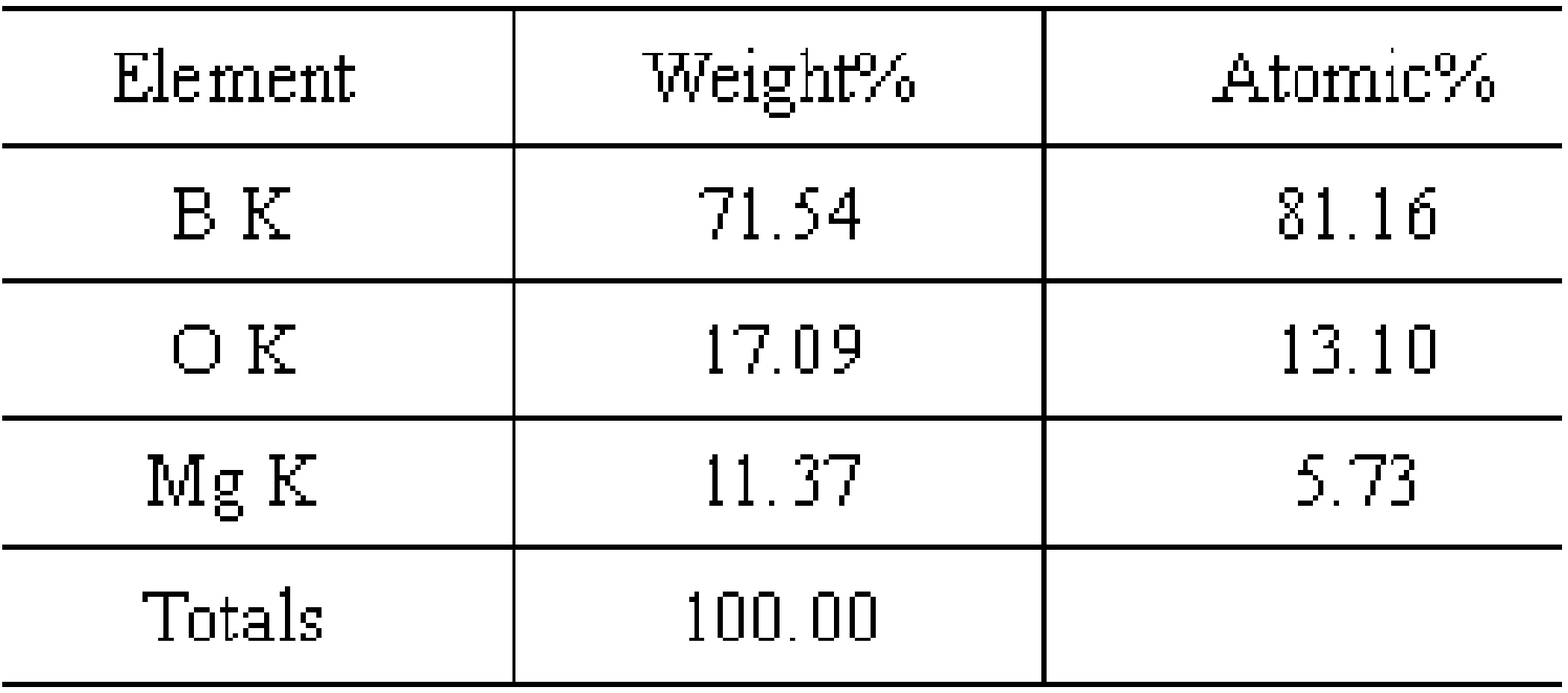}
\end{figure}

%%%%%%%%%%%%%%%%%%%%%%%%%%%%%%%%%%%%%%%%%%%%%%%%%%%%%%%%%%%%
\clearpage Fig.10
%\section{EDX Light second phase 0.5h}
%%%%%%%%%%%%%%%%%%%%%%%%  FIGURE 10 EDX Light second phase 0.5h  %%%%%%%%%%%%%%%%%%%%%%%%%
\begin{figure}[hp]
\centering
\includegraphics[width=180pt]{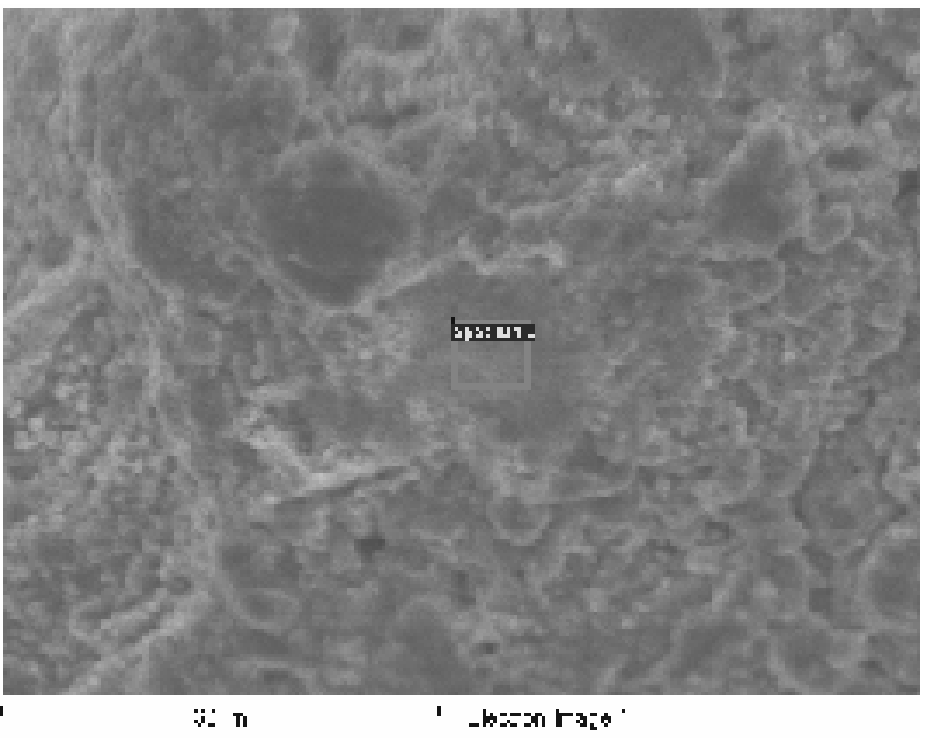}

\includegraphics[width=180pt]{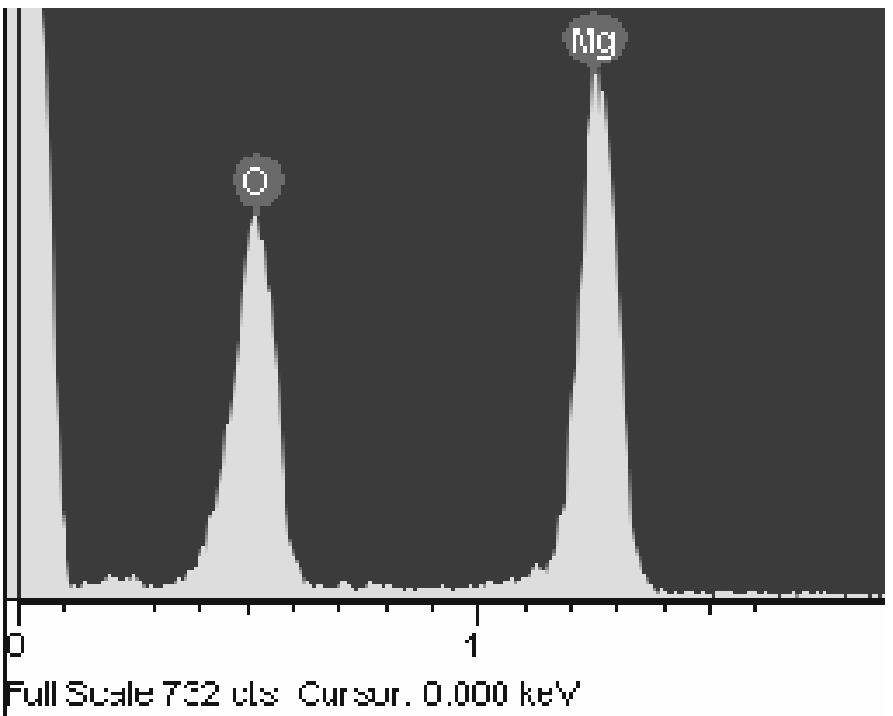}

\includegraphics[width=180pt]{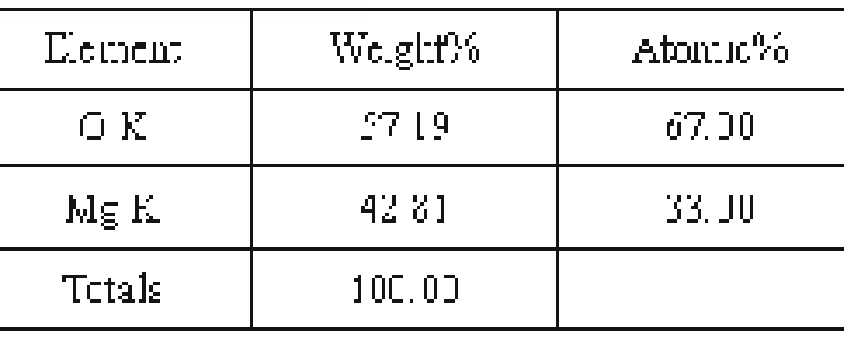}
\end{figure}

%%%%%%%%%%%%%%%%%%%%%%%%%%%%%%%%%%%%%%%%%%%%%%%%%%%%%%%%%%%%

\clearpage Fig.11
%\section{diff sintering time M-H}
%%%%%%%%%%%%%%%%%%%%%%%%  FIGURE 11 diff sintering time M-H  %%%%%%%%%%%%%%%%%%%%%%%%%
\begin{figure}[hp]
\centering
\includegraphics[width=300pt]{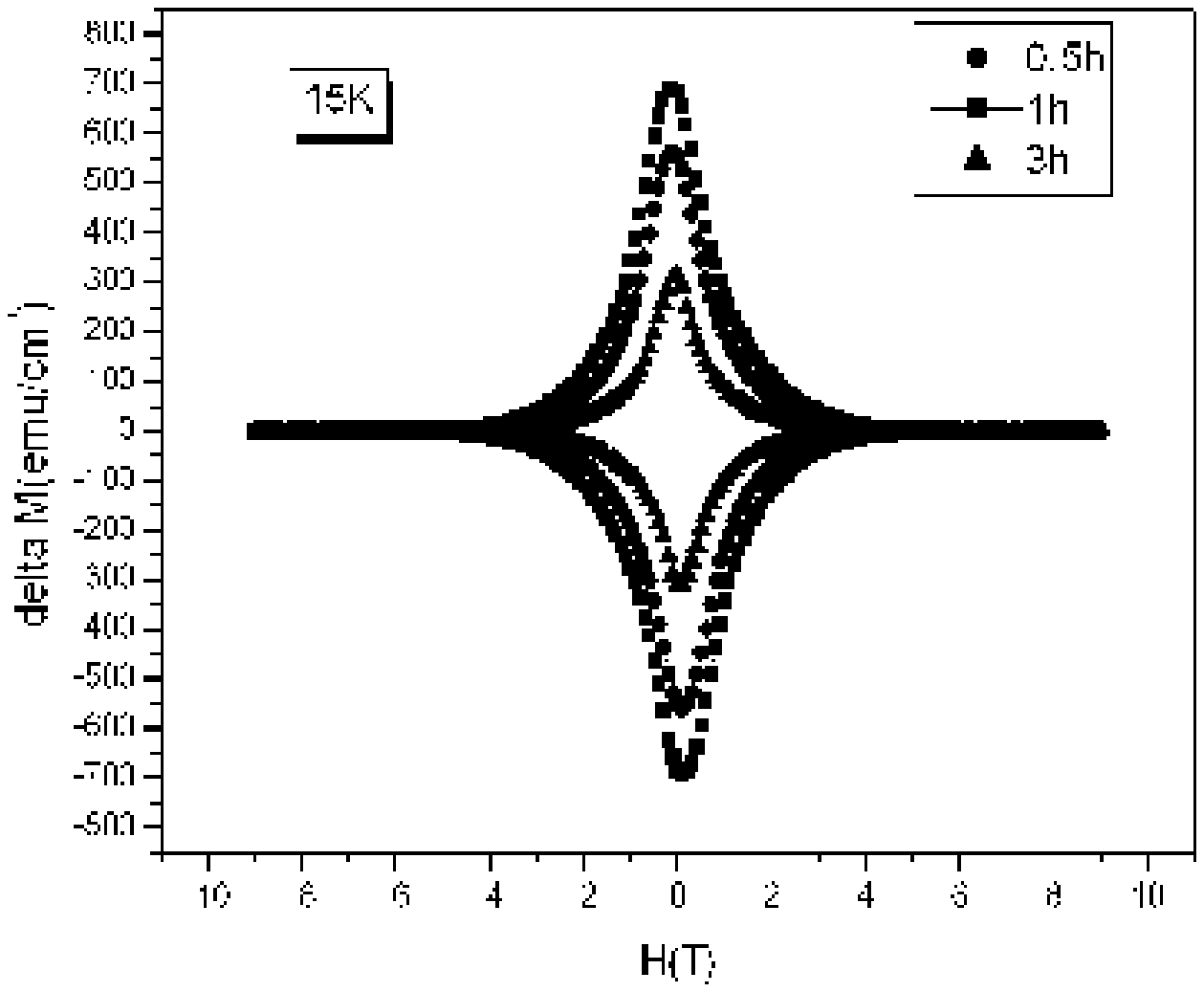}
\end{figure}

%%%%%%%%%%%%%%%%%%%%%%%%%%%%%%%%%%%%%%%%%%%%%%%%%%%%%%%%%%%%

\clearpage Fig.12
%\section{diff sintering time Jc-H}
%%%%%%%%%%%%%%%%%%%%%%%%  FIGURE 12 diff sintering time Jc-H  %%%%%%%%%%%%%%%%%%%%%%%%%
\begin{figure}[hp]
\centering
\includegraphics[width=300pt]{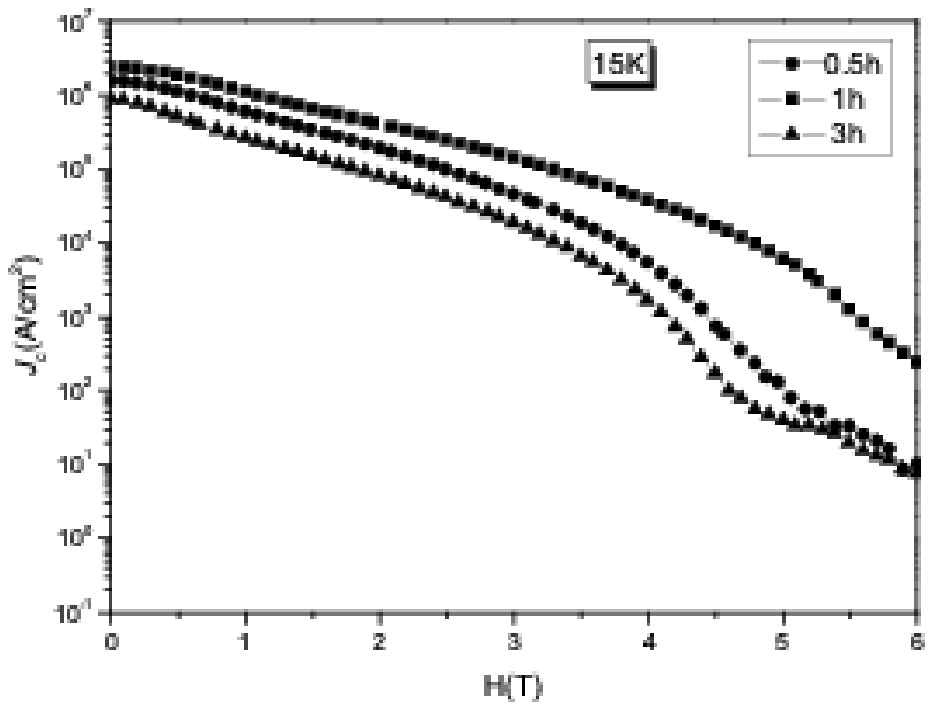}
\end{figure}

%%%%%%%%%%%%%%%%%%%%%%%%%%%%%%%%%%%%%%%%%%%%%%%%%%%%%%%%%%%%
\end{document}